\documentclass[12pt]{iopart}
\usepackage{iopams}

\usepackage{graphicx}
\usepackage{hyperref}
\usepackage{esint}
\usepackage{cite}
\usepackage{bm}

\def\bra#1{\langle #1 |}
\def\ket#1{| #1 \rangle}

\begin{document}

\title{Phase Diagram of Bipartite Entanglement}

\author{Paolo Facchi$^{1,2}$, Giorgio Parisi$^{3}$, Saverio Pascazio$^{1,2}$, Antonello Scardicchio$^{4,5}$, and Kazuya Yuasa$^6$}
\address{$^{1}$Dipartimento di Fisica and MECENAS, Universit\`a di Bari, I-70126  Bari, Italy}
\address{$^{2}$INFN, Sezione di Bari, I-70126 Bari, Italy}
\address{$^{3}$Dipartimento di Fisica, Sapienza Universit\`a di Roma, INFN, Sezione di Roma 1, and CNR-Nanotec, Rome Unit, I-00185 Rome, Italy}
\address{$^{4}$The Abdus Salam ICTP, I-34151 Trieste, Italy}
\address{$^{5}$INFN, Sezione di Trieste, I-34127 Trieste, Italy}
\address{$^{6}$Department of Physics, Waseda University, Tokyo 169-8555, Japan}

\date{\today}

\begin{abstract}
We investigate the features of the entanglement spectrum (distribution of the eigenvalues of the reduced density matrix) of a large quantum system in a pure state. We consider all R\'enyi entropies and recover purity and von Neumann entropy as particular cases.
We construct the phase diagram of the theory and unveil the presence of two critical lines.
\end{abstract}


\maketitle

\section{Introduction}
\label{sec:intro}

The notion of entanglement is central in the emerging fields of quantum technologies and quantum applications~\cite{NC}. Entanglement is a genuine non-classical feature of quantum states~\cite{BZbook} and characterizes the nonclassical correlations among the different components of a quantum system. 
It can be measured in terms of different quantities, such as purity and von Neumann entropy~\cite{woot,AFOV,H4}.

For large quantum systems, the distribution of bipartite entanglement is pivotal to understand the features of the many-body wave function. Interestingly, random pure states play a crucial role in this context: they are characterized by a large entanglement and display a number of interesting features. The first studies on this subject date back to forty years ago and focused on the average purity of a bipartite system, that turns out to be almost minimal for randomly sampled states~\cite{Lubkin}. These findings were extended to 
the average von Neumann entropy~\cite{Page,Pageproof1,Pageproof2,Pageproof3} and to higher moments~\cite{Giraud}, and are essentially  a consequence of the concentration of measure for the so-called entanglement spectrum (the eigenvalues of the reduced density matrix)~\cite{Winter}.

The typical entanglement spectrum was eventually determined~\cite{FMPPS,DPFPPS,DPFGPPS} and displayed the presence of phase transitions. These studies focused on the purity, and were soon extended to different R\'enyi entropies~\cite{majumdar,nadal} and eventually to the von Neumann entropy~\cite{vnentr}. It is in fact somewhat surprising that a number of interesting results can be obtained analytically, probably because they hinge on the Coulomb gas method~\cite{Dyson,forrester} and the ensuing saddle point equations~\cite{Fabio1,Fabio2,Fabio3,Fabio4}. 

In the approach proposed in Ref.~\cite{FMPPS} one ``biases" the amount of entanglement across the (given) bipartition and studies typicality constrained at such entanglement. This yields a family of entanglement spectra that depend on the adopted measure of entanglement and whose features are of great interest. 
One unveils the presence of two phase transitions, as the entanglement between the two partitions is changed. One of them is related to the ``evaporation" of the largest eigenvalue, which splits off from the continuous distribution of eigenvalues~\cite{Thouless1,Thouless2}, while the other one to the vanishing of the smallest eigenvalue and to the squeezing of the distribution against the hard wall at zero (pushed-to-pulled transition)~\cite{Witten,tHooft,BIPZ,Majumdar2014,Fabio5}.

In this Article we shall scrutinize the features of these phase transitions for all R\'enyi entropies. We shall find that, as anticipated in~\cite{vnentr}, the phase transitions for the von Neumann entropy are smoother than for all R\'enyi entropies, and in particular one of them becomes continuous when the other one is of first order.

This work is organized as follows.
We define the problem and set up our notation in Section~\ref{sec:setup}.
The \emph{entangled}, \emph{typical} and \emph{separable phases} are analyzed in Secs.~\ref{sec:entph},~\ref{sec:typph} and~\ref{sec:sepph}, respectively.
The phase diagram is drawn in Sec.~\ref{sec:phtrans}. We conclude in Sec.~\ref{sec:concl}.

\section{Setting up the problem}
\label{sec:setup}

Consider a bipartite system in the Hilbert space $\mathcal{H}=\mathcal{H}_A\otimes \mathcal{H}_{\bar{A}}$, described by a pure state $\ket{\psi}$. 
The reduced density matrix of subsystem $A$,
\begin{equation}
\varrho_A=\tr_{\bar{A}}\ket{\psi}\bra{\psi},
\label{eq:aabar}
\end{equation}
is a (Hermitian) positive matrix of unit trace, $\tr\varrho_A=1$.
We quantify the bipartite entanglement between $A$ and $\bar{A}$ by the R\'enyi entropy of $\varrho_A$,
\begin{equation}
S_q(\vec{\lambda})
=-\frac{1}{q-1}
\ln\!\left(
\sum_{k=1}^N \lambda_k^q
\right),
\label{eq:entropy}
\end{equation}
where $N=\dim\mathcal{H}_A$, $\vec{\lambda}=(\lambda_1,\dots,\lambda_N)\in \Delta_{N-1}$ are  the eigenvalues (Schmidt coefficients) of $\varrho_A$, and $\Delta_{N-1}$ is the 
simplex of eigenvalues ($\lambda_k\ge0$, $\sum_k\lambda_k=1$). 
We are interested in balanced bipartitions: $N=\dim\mathcal{H}_A=\dim\mathcal{H}_{\bar{A}}$. 
The R\'enyi entropy ranges $0\leq S_q \leq \ln N$, where the minimum and maximum values are obtained, respectively, for separable and maximally entangled vector states $\ket{\psi}$.
Note that the R\'enyi entropy reduces to the von Neumann entropy $S_1=-\sum_k\lambda_k\ln\lambda_k$ in the limit $q\to1$.

We will focus on the typical features of the afore-mentioned eigenvalues. For random states $\ket{\psi}$, uniformly sampled on the unit sphere $\bra{\psi}\psi\rangle=1$, 
the eigenvalues $\vec{\lambda}$ are distributed according to the (Haar) joint probability density function~\cite{LLoydPagels,ZS,IZ} 
\begin{equation}
p_{N}(\vec{\lambda})=C_N \prod_{1\leq j<k\leq N}{(\lambda_j-\lambda_k)^2} ,
\label{eq:Haar_invariant}
\end{equation}
$C_N$ being a normalization factor. For large $N$, the distribution $p_N$ concentrates around a typical set $\vec{\lambda}$, that maximizes $p_N$~\cite{Winter}, and the typical spectral distribution of $\varrho_A$ follows a Mar\v{c}enko-Pastur law~\cite{MP} with support $[0,4/N]$~\cite{vnentr}.

We shall ask here how the entanglement spectrum is distributed in a system with a given amount of bipartite entanglement, conditioned at a given value of the entropy $S_q$. This is nothing but a constrained maximization problem.
Let
\begin{equation}
u=\ln N-S_q(\vec{\lambda})=\frac{1}{q-1}\ln\!\left(\frac{1}{N}\sum_k(N\lambda_k)^q\right),
\label{eq:uuu}
\end{equation}
which quantifies the deviation of the entropy $S_q$ from its maximum value $\ln N$.
Then, given a value $u \in [0, \ln N]$, we seek  $\vec{\lambda}_{\mathrm{max}}$ such that 
\begin{equation}
\label{eq:constrmax}
p_N(\vec{\lambda}_{\mathrm{max}})=\max\left\{ p_N(\vec{\lambda}) \,:\,  \vec{\lambda} \in \Delta_{N-1}, \,  S_q(\vec{\lambda}) = \ln N - u \right\}.
\end{equation}
Introducing two Lagrange multipliers $\xi$ and $\beta$, that constrain the eigenvalue normalization and the deviation $u$ of the entropy from its maximum $\ln N$, respectively, the problem is translated into the (unconstrained) minimization of the potential
\begin{eqnarray}
V(\vec{\lambda},\xi,\beta)
&=&{-\frac{2}{N^2}}\sum_{j<k}\ln |\lambda_j-\lambda_k|
+\xi\left(
\sum_k\lambda_k- 1
\right)
\nonumber\\
& &+\beta
\left[
\frac{1}{q-1}\ln\!\left(
\frac{1}{N}\sum_k(N\lambda_k)^q
-\sum_k\lambda_k+1
\right)
-u
\right]
\label{eq:potential}
\end{eqnarray}
with respect to $\vec{\lambda}$, $\xi$, and $\beta$. 
Note that we have added two terms in the logarithm that expresses the constraint on the entropy; otherwise, the expression is not well defined for $q\to1$ before imposing the other constraint on the eigenvalue normalization.  
The potential $V$ can be viewed as the energy of a gas of point charges (eigenvalues) distributed in the interval $[0,1]$ with a 2D  (logarithmic) Coulomb repulsion, subject to two external electric fields proportional to $\xi$ and $\beta$. The logarithmic form of the interaction is a direct consequence of the product form~(\ref{eq:Haar_invariant}) of the joint probability density.

It is worth noting that this problem can be equivalently framed in the statistical mechanics of points on the simplex  $\Delta_{N-1}$ with  partition function~\cite{vnentr} 
\begin{equation}
Z_N=  \int_{\Delta_{N-1}}  \e^{-\beta N^2 E(\vec{\lambda})} 
p_N(\vec{\lambda})\, \rmd^N\lambda ,
\label{eq:partitionfunction}
\end{equation}
with an ``energy density'' $E(\vec{\lambda})= 
\ln N - S_q(\vec{\lambda})$ and an inverse ``temperature'' $\beta$. In the thermodynamic limit $N\to\infty$, one looks at the maximum of the integrand, that is at the minimum of the potential (\ref{eq:potential}).
Large values of $\beta$ yield highly entangled  states, while  $\beta=0$ yields random states.

The saddle-point equations read
\begin{eqnarray}
\fl \qquad\quad \frac{\partial V}{\partial\lambda_j}
=
\frac{2}{N}
\sum_{k\neq j}
\frac{1}{N\lambda_k- N\lambda_j}
+
\frac{
\beta
}{q-1}
\frac{
q(N\lambda_j)^{q-1}
-
1
}{
\displaystyle
\frac{1}{N}\sum_k(N\lambda_k)^q
- \frac{1}{N}\sum_k N \lambda_k
+1
}
+\xi=0,
\label{eqn:SaddlePointLambda}
\\
\fl \qquad\quad \frac{\partial V}{\partial\beta}
=
\frac{1}{q-1}\ln\!\left(
\frac{1}{N}\sum_k(N\lambda_k)^q
-\frac{1}{N} \sum_k N \lambda_k
+1
\right)-u=0,
\label{eqn:SaddlePointBeta}
\\
\fl \qquad\quad \displaystyle
\frac{\partial V}{\partial\xi}
=
\frac{1}{N} \sum_k N \lambda_k-1
=0.
\label{eqn:SaddlePointXi}
\end{eqnarray}
When all the eigenvalues $\lambda_k$ are of order $O(1/N)$, we introduce the empirical eigenvalue distribution
\begin{equation}
\sigma(\lambda)=\frac{1}{N}\sum_k\delta(\lambda-N\lambda_k),
\end{equation}
with
\begin{equation}
\label{eq:normalisation}
\int\rmd\lambda\,\sigma(\lambda)=1,
\end{equation}
and Eqs.~(\ref{eqn:SaddlePointLambda})--(\ref{eqn:SaddlePointXi}) read
\begin{eqnarray}
2 \fint\rmd\lambda'\,
\frac{
\sigma(\lambda')
}{\lambda'-\lambda}
+
\beta
\rme^{-(q-1)u}
\frac{
q\lambda^{q-1}
-
1
}{q-1}
+\xi=0,
\label{eqn:SaddlePoint}
\\
\int\rmd\lambda\,\sigma(\lambda)\lambda=1,
\label{eqn:SigmaNormalization}
\\
u
=
\frac{1}{q-1}\ln\!\left(
\int\rmd\lambda\,\sigma(\lambda)\lambda^q
\right),
\label{eqn:Usigma}
\end{eqnarray}
with $\lambda=N\lambda_j$, and $\fint$ denoting the Cauchy principal value.

The above expressions are suitable to the limit $N\to+\infty$. The integral equation (\ref{eqn:SaddlePoint}) admits a solution $\sigma(\lambda)$ that lies within a compact support 
\begin{equation}
\lambda \in [a, b], \qquad  \mathrm{with} \quad a \geq 0 ,
\end{equation} 
and can be obtained via a theorem by Tricomi~\cite{Tricomi}.
Let us change the variable from $\lambda$ to $x\in[-1,1]$ by
\begin{equation}
\lambda=\delta (x + \alpha),
\qquad
\delta=\frac{b-a}{2}, \qquad
\alpha =\frac{b+a}{b-a} 
= 1 + \frac{a}{\delta},
\label{eq:deltaalphadef}
\end{equation}
so that the distribution $\phi(x)$ of $x$ is related to $\sigma(\lambda)$ through
\begin{equation}
\sigma(\lambda)=\frac{1}{\delta}\phi(x).
\end{equation}
In terms of these quantities, the equations in (\ref{eqn:SaddlePoint})--(\ref{eqn:SigmaNormalization}) read
\begin{eqnarray}
\frac{1}{\pi}
\fint_{-1}^1\rmd y\,
\frac{
\phi(y)
}{y-x}
=
-
\frac{\delta}{2\pi}
\left(
\beta
\rme^{-(q-1)u}
\frac{
q \delta^{q-1} (x+\alpha)^{q-1}
-
1
}{q-1}
+\xi
\right),
\\
\int_{-1}^1\rmd x\,\phi(x)x=\frac{1}{\delta} -\alpha.
\label{eqn:TricomiVN}
\end{eqnarray}
Their solution is given by
\begin{eqnarray}
\phi(x)
=\frac{1}{\pi\sqrt{1-x^2}}
[
1-Ax
+
B
h(x,\alpha)
],
\label{eqn:ConstraintNormalization}
\end{eqnarray}
where
\begin{eqnarray}
A=
\frac{\delta}{2}\left(
\beta
\rme^{-(q-1)u}
\frac{
q\delta^{q-1}
-
1
}{q-1}
+\xi
\right),
\qquad 
B= \frac{1}{2}
\beta
q\delta^q
\rme^{-(q-1)u}
,
\label{eqn:AB}
\\
h(x,\alpha)
=
\frac{1}{\pi}
\fint_{-1}^1\rmd y\,\frac{\sqrt{1-y^2}}{y-x}
\frac{
(y+\alpha)^{q-1}
-
1
}{q-1}.
\end{eqnarray}
The last equation (\ref{eqn:Usigma}) reads
\begin{equation}
u
=
\frac{1}{q-1}\ln\!\left(
\delta^q
\int_{-1}^1\rmd x\,\phi(x)(x+\alpha)^q
\right).
\label{eqn:u_in_x}
\end{equation}
We are now ready to investigate the behavior of the entanglement spectrum as $q$ and $u$ are varied.
Remember that, from Eq.\ (\ref{eq:uuu}), $u$ can be viewed as the opposite of the entanglement between the two bipartitions of the total system. See Eqs.\ 
(\ref{eq:aabar})--(\ref{eq:entropy}).

\section{Entangled phase ($\alpha>1$, small $u$)}
\label{sec:entph}

Large values of bipartite entanglement correspond to  small values of $u$. They are obtained for low temperatures [large $\beta$ in Eq.\ (\ref{eq:partitionfunction})].  The limit of the empirical measure is compactly supported in $\lambda \in [a,b]$, with
\begin{equation}
0<a< b, \qquad \Rightarrow \quad \alpha>1.
\end{equation}
The values of the extremes of the support of the distribution, $a$ and $b$, and thus $\delta$ and $\alpha$ in~(\ref{eq:deltaalphadef}), can be determined by imposing the constraint (\ref{eqn:TricomiVN}) and the conditions of regularity at both ends of the distribution, 
\begin{equation}
\phi(-1)=0, \qquad \mathrm{and} \quad \phi(1)=0,
\end{equation} 
which yield
\begin{eqnarray}
-\frac{1}{2}A
-
B
g(\alpha)
=\frac{1}{\delta}-\alpha,
\label{eqn:Cond1}
\\
1-A
+
B
h(1,\alpha)
=0,
\vphantom{\frac{1}{\delta}}
\label{eqn:Cond2}
\\
1+A
+
B
h(-1,\alpha)
=0,
\vphantom{\frac{1}{\delta}}
\label{eqn:Cond3}
\end{eqnarray}
with
\begin{equation}
g(\alpha)
=
\frac{1}{\pi}
\int_{-1}^1\rmd y\,\sqrt{1-y^2}
\frac{
(y+\alpha)^{q-1}
-
1
}{q-1}.
\end{equation}
One gets
\begin{eqnarray}
A=-\frac{h(1,\alpha)-h(-1,\alpha)}{h(1,\alpha)+h(-1,\alpha)},
\\
B=-\frac{2}{h(1,\alpha)+h(-1,\alpha)},
\\
\delta
=\left(
\alpha
-\frac{1}{2}A
-Bg(\alpha)
\right)^{-1},
\end{eqnarray}
for $\alpha>1$.
These expressions are valid for what we shall call the ``entangled" phase. There are two interesting particular cases:
\begin{itemize}
\item\textbf{von Neumann entropy, $\bm{q\to1}$:}
In this limit, we have
\begin{eqnarray}
A=(\alpha+\sqrt{\alpha^2-1})\ln[2(\alpha-\sqrt{\alpha^2-1})],
\vphantom{\frac{4}{3\alpha+\sqrt{\alpha^2-1}}}
\\
B=\alpha+\sqrt{\alpha^2-1},
\vphantom{\frac{4}{3\alpha+\sqrt{\alpha^2-1}}}
\\
\delta
=\frac{
4
}{
3\alpha
+\sqrt{\alpha^2-1}
},
\end{eqnarray}
for $\alpha>1$.
Finally, $u$ is given by~\cite{vnentr}
\begin{equation}
u=\ln\!\left(1-\frac{1}{2\beta}\right)+\frac{1}{\beta},
\label{eq:ubeta}
\end{equation}
where $\beta$ is obtained  from the definition of $B$ in (\ref{eqn:AB}):
\begin{equation}
\beta
=
\frac{1}{2}
(
\alpha+\sqrt{\alpha^2-1}
)
(
3\alpha
+\sqrt{\alpha^2-1}
).
\end{equation}
The function $\beta(\alpha)$ is strictly increasing for $\alpha>1$, with  $\beta(1)=3/2$ and $\beta\to\infty$ as $\alpha\to\infty$, while $u(\beta)$ is strictly decreasing for $\beta>3/2$ with $u(3/2) = 2/3 +\ln(2/3)$ and $u\to 0$ as $\beta\to\infty$. Therefore Eq.~(\ref{eq:ubeta}) can be inverted to get $\alpha$ as a function of~$u$.

\item\textbf{purity, $\bm{q=2}$:}
In this case, we have
\begin{equation}
A=-2(\alpha-1),\qquad
B=2,\qquad
\delta
=\alpha^{-1}
\qquad(\alpha>1),
\end{equation}
and 
\begin{equation}
\phi(x)=\frac{2}{\pi}\sqrt{1-x^2},\quad
\sigma(\lambda)
=\frac{2\alpha^2}{\pi}\sqrt{ \frac{1}{\alpha^2} -(\lambda -1)^2 },
\end{equation}
with
\begin{equation}
\alpha = \frac{1}{2}(\mathrm{e}^u -1)^{-1/2}.
\end{equation}

\end{itemize}

As $u$  (and thus the temperature) is increased, and $\alpha\downarrow1$ accordingly, the left end $a$ of the distribution $\sigma(\lambda)$ of the eigenvalues touches the boundary at $\lambda=0$,  that is $a=0$ ($\alpha=1$), and the system reaches the first phase-transition line $u_C(q)$, where 
\begin{eqnarray}
A_C(q)
=
-
1
-
\frac{1}{q-1}
\left(
1
-
\frac{
\sqrt{\pi}\Gamma(q+1)
}{
2^{q-1}\Gamma(q-1/2)
}
\right),
\\
B_C(q)
=
\frac{
\sqrt{\pi}\Gamma(q+1)
}{
2^{q-1}\Gamma(q-1/2)
},
\\
\delta_C(q)
=
\frac{2(q+1)}{3q},
\label{eq:deltaCq}
\end{eqnarray}
at $\alpha=1$. The explicit expression of $u$ along the critical line is
\begin{equation}
u_C(q)=\frac{1}{q-1}\ln\!\left[
\left(
\frac{4(q+1)}{3q}
\right)^q
\frac{\Gamma(q+3/2)}{\sqrt{\pi}\Gamma(q+2)}
\right],
\label{eq:uCq}
\end{equation}
and is obtained  by evaluating at $\delta = \delta_C(q)$ the expression of $u$ derived in~(\ref{eq:utypicalphase})  in the next section. The (whole) phase diagram will be shown in Fig.~\ref{fig:phasediagram}.

\section{Typical phase ($\alpha=1$, intermediate $u$)}
\label{sec:typph}

We keep increasing $u$ (and temperature), and thus lowering entanglement, beyond the first critical line $u_C(q)$.
In this regime, 
\begin{equation}
0=a< b, \qquad \Rightarrow \quad \alpha=1,
\end{equation}
and we impose only the constraint (\ref{eqn:TricomiVN}) and the condition 
\begin{equation}
\phi(1)=0,
\end{equation}
namely the conditions (\ref{eqn:Cond1}) and (\ref{eqn:Cond2}).
We get
\begin{eqnarray}
A
=
1
-
\frac{
2/\delta-1
}{q-1}
\left(
q+1
-
\frac{\sqrt{\pi}\Gamma(q+2)}{2^q\Gamma(q+1/2)}
\right)
,
\\
B
=
(
2/\delta
-1
)
\frac{\sqrt{\pi}\Gamma(q+2)}{2^q\Gamma(q+1/2)}
.
\end{eqnarray}
An explicit expression of $u$ in this phase is available as a function of $\delta$ and $q$,
\begin{equation}
u=\frac{1}{q-1}
\ln\!\left[
\left(
\frac{q+1}{\delta}
-\frac{q-1}{2}
\right)
\frac{(2\delta)^q\Gamma(q+1/2)}{\sqrt{\pi}\Gamma(q+2)}
\right].
\label{eq:utypicalphase}
\end{equation}

The first critical $u_C(q)$ line is reached at $\delta = \delta_C(q)$ in~(\ref{eq:deltaCq}).
A second critical line $u_E(q)$ is reached at 
\begin{equation}
\delta_E(q)=2, \qquad  (\beta_E=0),
\end{equation} 
where
\begin{equation}
u_E(q)=\frac{1}{q-1}\ln\!\left(
\frac{2^{2q}\Gamma(q+1/2)}{\sqrt{\pi}\Gamma(q+2)}
\right),
\end{equation}
and 
\begin{equation}
\phi_{\mathrm{MP}}(x)
=\frac{1}{\pi}
\sqrt{\frac{1-x}{1+x}},\qquad
\sigma_{\mathrm{MP}}(\lambda)
=\frac{1}{2\pi}
\sqrt{\frac{4-\lambda}{\lambda}}.
\end{equation}
This is the Mar\v{c}enko-Pastur law, the distribution of typical states.

\section{Separable phase (large $u$)}
\label{sec:sepph}

We can reach lower values of entanglement, towards the separable states, by increasing $u$ above the second critical line $u_E(q)$. This corresponds to negative temperatures $\beta<0$ of the statistical-mechanics model.

In the separable phase, one eigenvalue $\lambda_1=\mu=O(1)$ while the others $\lambda_k=O(1/N)$ ($k\ge2$).
In this case, the saddle point equations in (\ref{eqn:SaddlePointLambda})--(\ref{eqn:SaddlePointXi})  reduce, for large $N$, to
\begin{eqnarray}
\frac{2}{N^2}
\sum_{k\ge2,k\neq j}
\frac{1}{\lambda_k-\lambda_j}
+\xi=0
\qquad(j\ge2),
\label{eqn:SaddlePointLambdaMu2N}
\\
\sum_{k\ge2}\lambda_k
=1-\mu,
\qquad
\xi
=
-\beta
\frac{
q(N\mu)^{q-1}
-
1
}{
(q-1)N^{q-1}\mu^q
},
\label{eqn:SaddlePointLambdaMu1N}
\\
u
=
\frac{1}{q-1}\ln(
N^{q-1}\mu^q
-\mu
+1
).
\label{eqn:SaddlePointBetaMuN}
\label{eqn:SaddlePointXiMuN}
\end{eqnarray}
By introducing the empirical distribution
\begin{equation}
\tilde{\sigma}(\lambda)
=\frac{1}{N-1}\sum_{k\ge2}\delta\!\left(
\lambda
-
\frac{N-1}{1-\mu}\lambda_k
\right),
\end{equation}
these equations become
\begin{eqnarray}
2\fint\rmd\lambda'\,
\frac{
\tilde{\sigma}(\lambda')
}{\lambda'-\lambda}
+\xi(1-\mu)=0,
\label{eqn:SaddlePointLambdaMu2NTricomi}
\\
\int\rmd\lambda\,
\tilde{\sigma}(\lambda)
=1,\qquad
\int\rmd\lambda\,
\tilde{\sigma}(\lambda)
\lambda
=1,
\label{eqn:MuNormalizations}
\end{eqnarray}
with
\begin{eqnarray}
u
=
\frac{1}{q-1}\ln(
N^{q-1}\mu^q
-\mu
+1
),
\\
\xi
=
-\beta
\frac{
q(N\mu)^{q-1}
-
1
}{
(q-1)N^{q-1}\mu^q
},
\end{eqnarray}
and $\lambda=(N-1)\lambda_j/(1-\mu)$.
The set of equations (\ref{eqn:SaddlePointLambdaMu2NTricomi})--(\ref{eqn:MuNormalizations}) is formally equivalent to (\ref{eqn:SaddlePoint})--(\ref{eqn:SigmaNormalization}) with $\beta\to0$ and $\xi\to\xi(1-\mu)$.
Therefore, the solution is obtained by translating the result for $\beta=0$ in the previous subsection,
\begin{equation}
\tilde{\sigma}_{\mathrm{MP}}(\lambda)
=\frac{1}{2\pi}\sqrt{\frac{4-\lambda}{\lambda}},
\qquad
\xi
=
-\beta
\frac{
q(N\mu)^{q-1}
-
1
}{
(q-1)N^{q-1}\mu^q
}
=\frac{1}{1-\mu}.
\end{equation}

\begin{figure}[t]
\includegraphics[width=0.5\textwidth]{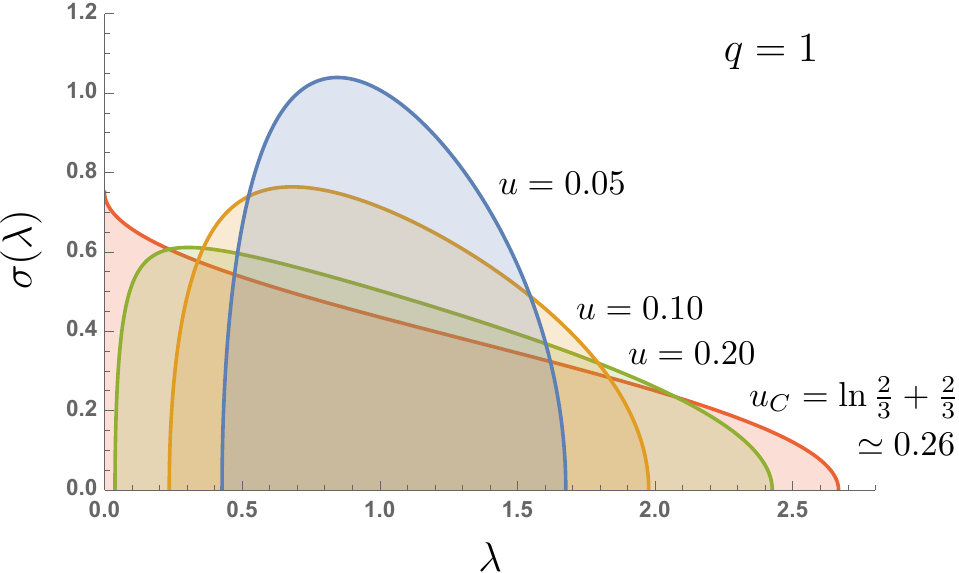}
\includegraphics[width=0.5\textwidth]{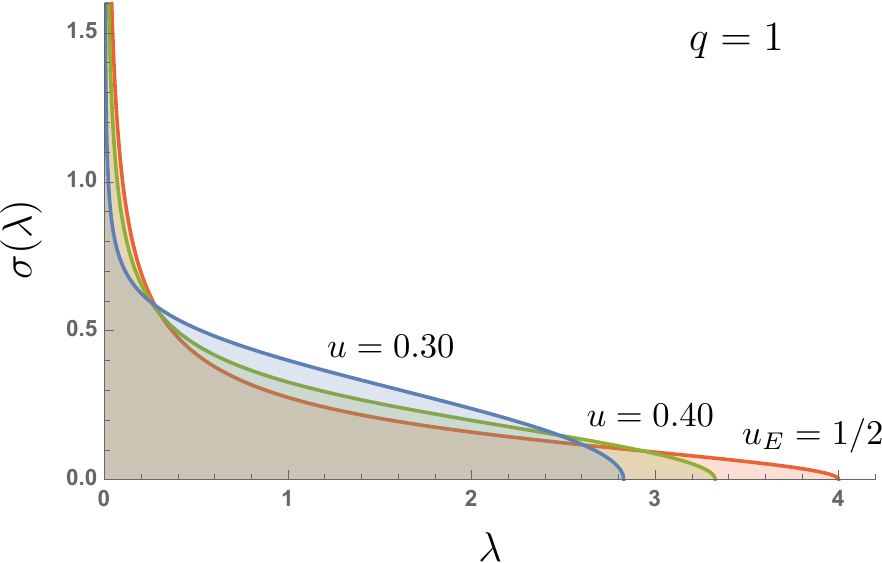}
\caption{(Color online) Entanglement spectra $\sigma(\lambda)$ at $q=1$ for various values of the von Neumann entropy  $S_{1}= \ln N - u$: entanglement decreases as $u$ increases.
(a) Entangled phase: $0\leq u \leq u_C(1) = \ln (2/3) + 2/3$,
(b) Typical phase: $u_C(1)\leq u \leq u_E(1)=1/2$. Both $u_C(1)$ and $u_E(1)$ are critical values belonging to the critical lines $u_C(q)$ and $u_E(q)$, see Fig.~\ref{fig:phasediagram}.
}
\label{fig:figeigvaldistr1}
\end{figure}

\begin{figure}[t]
\includegraphics[width=0.5\textwidth]{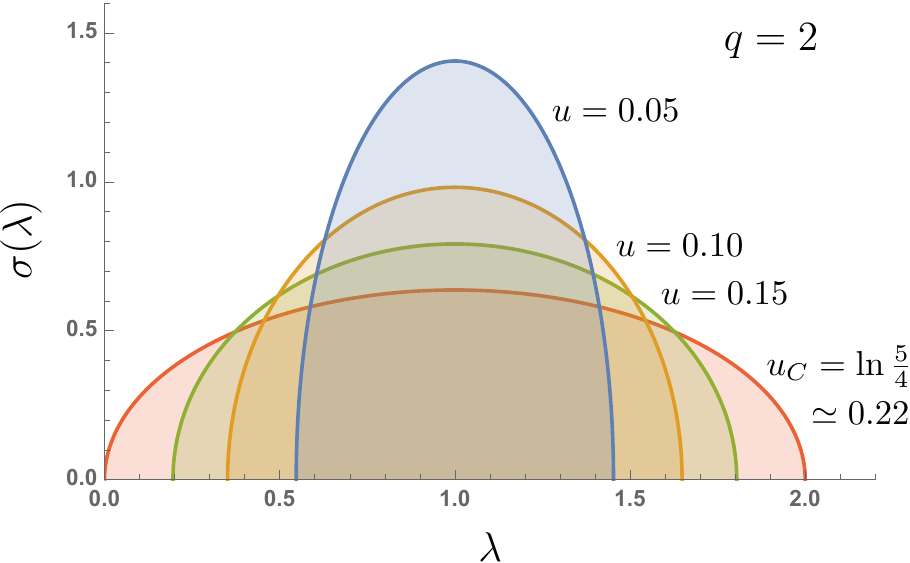}
\includegraphics[width=0.5\textwidth]{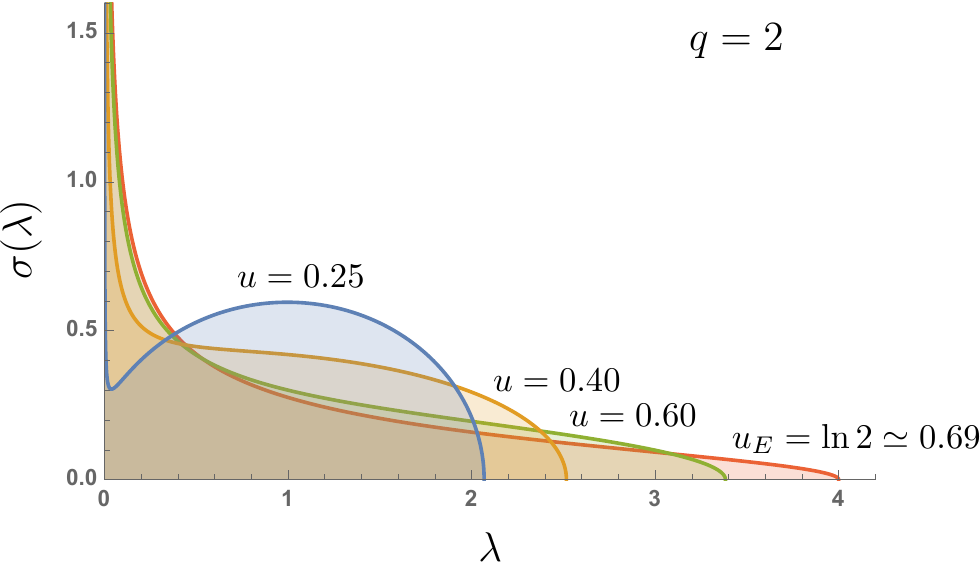}
\caption{(Color online) Entanglement spectra $\sigma(\lambda)$ at $q=2$ for various values of (the logarithm of) purity  $S_{2}= \ln N - u$: entanglement decreases as $u$ increases.
(a) Entangled phase: $0\leq u \leq u_C(2) =\ln (5/4)$,
(b) Typical phase: $u_C(2) \leq u \leq u_E(2)=\ln 2$.
Both $u_C(2)$ and $u_E(2)$ are critical values belonging to the critical lines $u_C(q)$ and $u_E(q)$, see Fig.~\ref{fig:phasediagram}.
}
\label{fig:figeigvaldistr2}
\end{figure}

\begin{figure}[t]
\includegraphics[width=0.5\textwidth]{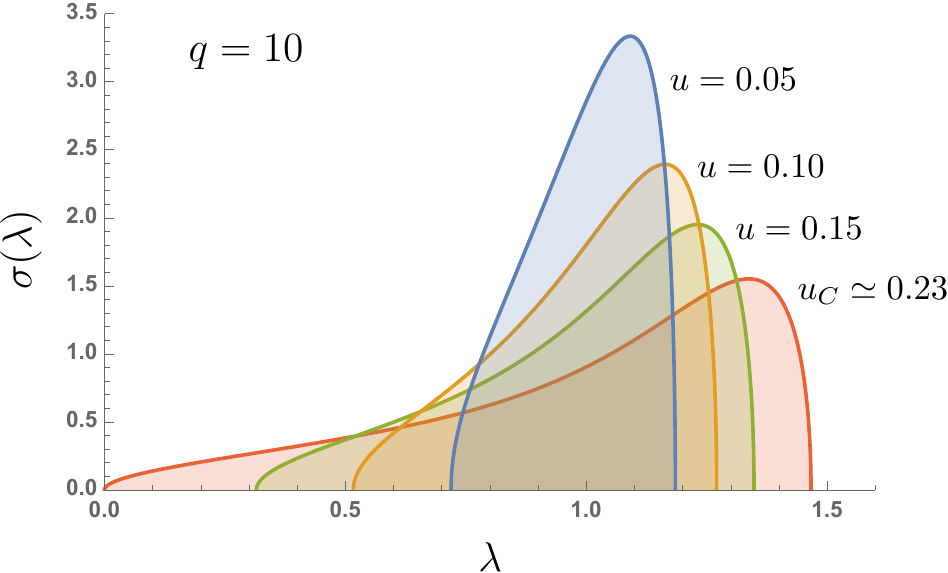}
\includegraphics[width=0.5\textwidth]{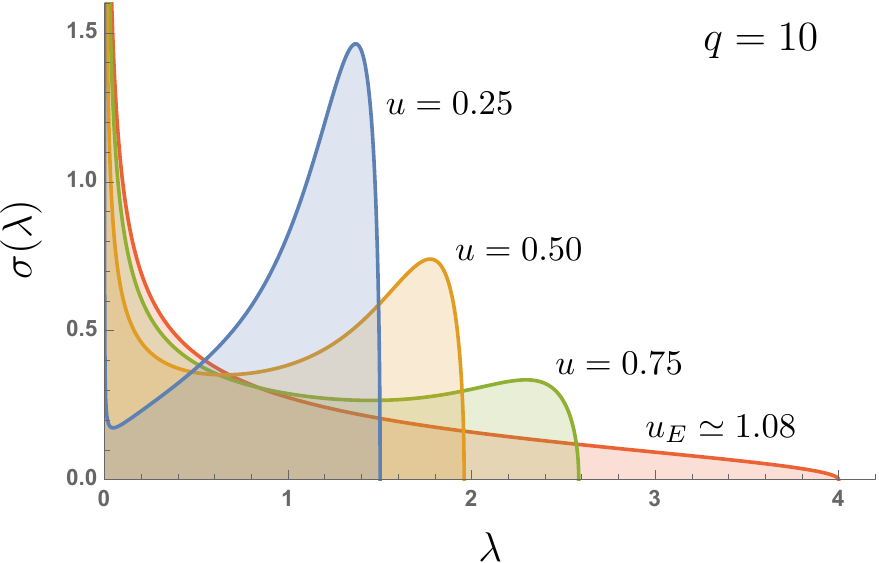}
\caption{(Color online) Entanglement spectra $\sigma(\lambda)$ at $q=10$ for various values of the R\'enyi entropy  $S_{10}= \ln N - u$: entanglement decreases as $u$ increases.
(a) Entangled phase: $0\leq u \leq u_C(10) \simeq 0.23 $,
(b) Typical phase: $u_C(10) \leq u \leq u_E(10) \simeq 1.08$.
Both $u_C(10)$ and $u_E(10)$ are critical values belonging to the critical lines $u_C(q)$ and $u_E(q)$, see Fig.~\ref{fig:phasediagram}.
 }
\label{fig:figeigvaldistr3}
\end{figure}

\begin{figure}[t]
\begin{center}
\includegraphics[width=0.5\textwidth]{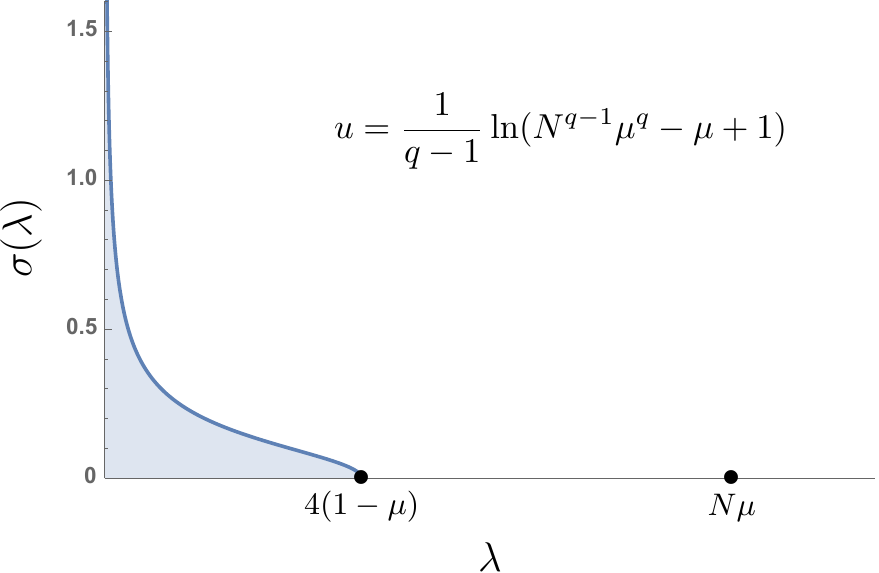}
\end{center}
\caption{(Color online) Entanglement spectra $\sigma(\lambda)$ in the separable phase $u_E(q) < u \leq \ln N$. One eigenvalue has evaporated from the spectrum sea, which has a Mar\v{c}enko-Pastur distribution.
}
\label{fig:figeigvaldistr4}
\end{figure}

\section{The phase diagram}
\label{sec:phtrans}

Some entanglement spectra are displayed in Figs.~\ref{fig:figeigvaldistr1}--\ref{fig:figeigvaldistr4}.
The spectra for $q=2$ and $q=1$ were shown in Ref.~\cite{DPFPPS} and~\cite{vnentr}, respectively. The spectra for a generic $q\neq 1,2$ are novel and we show an example in Fig.~\ref{fig:figeigvaldistr3}. The deformations of the spectra for different $q$ are very interesting and are easily understood by observing that large values of $q$ tend to attribute more ``weight" to large eigenvalues.
For any $q$, at $u=u_E(q)$, one eigenvalue evaporates from the spectrum sea $O(1/N)$ and becomes $O(1)$. See Fig.~\ref{fig:figeigvaldistr4}.

\begin{figure}[h]
\includegraphics[width=0.5\textwidth]{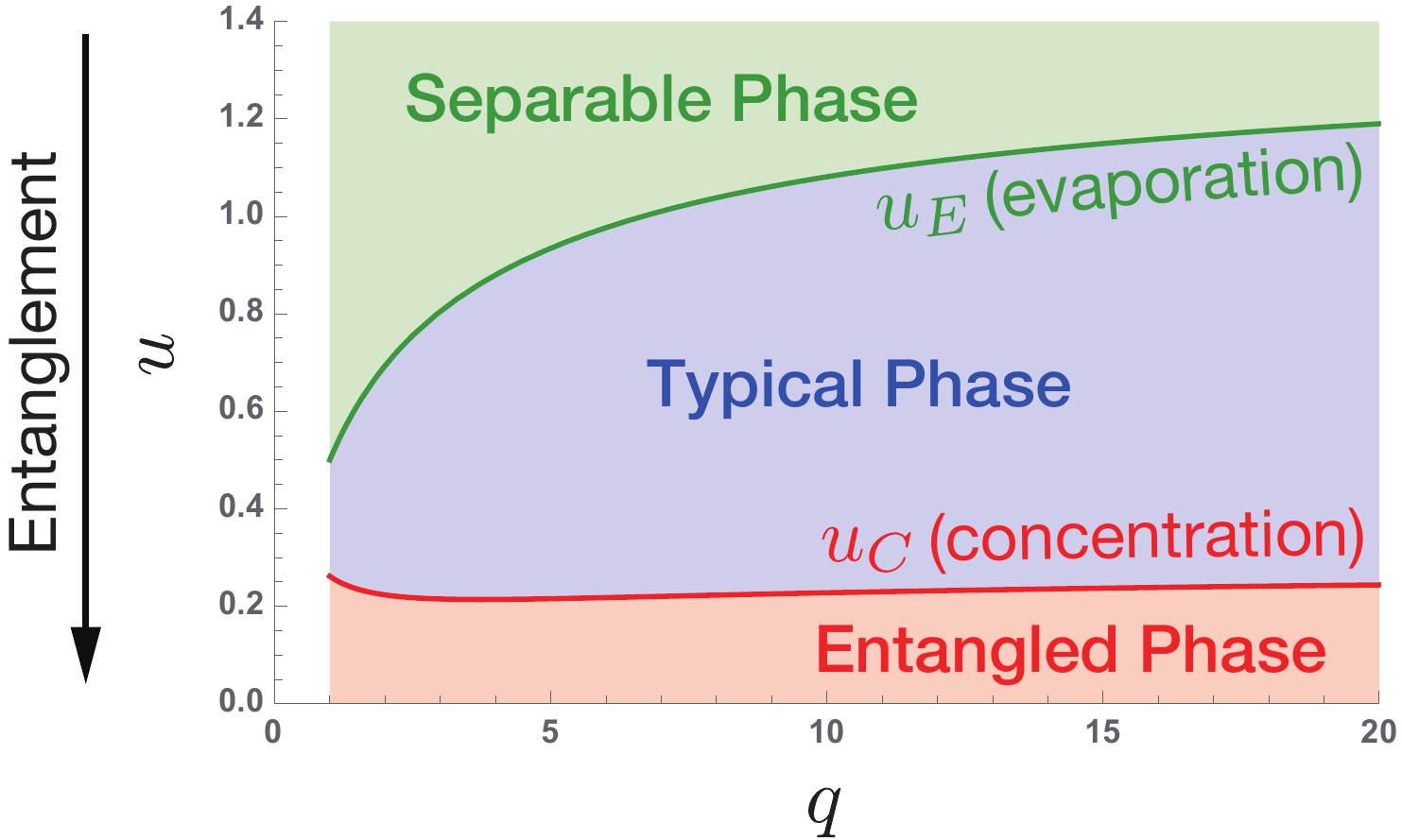}
\includegraphics[width=0.5\textwidth]{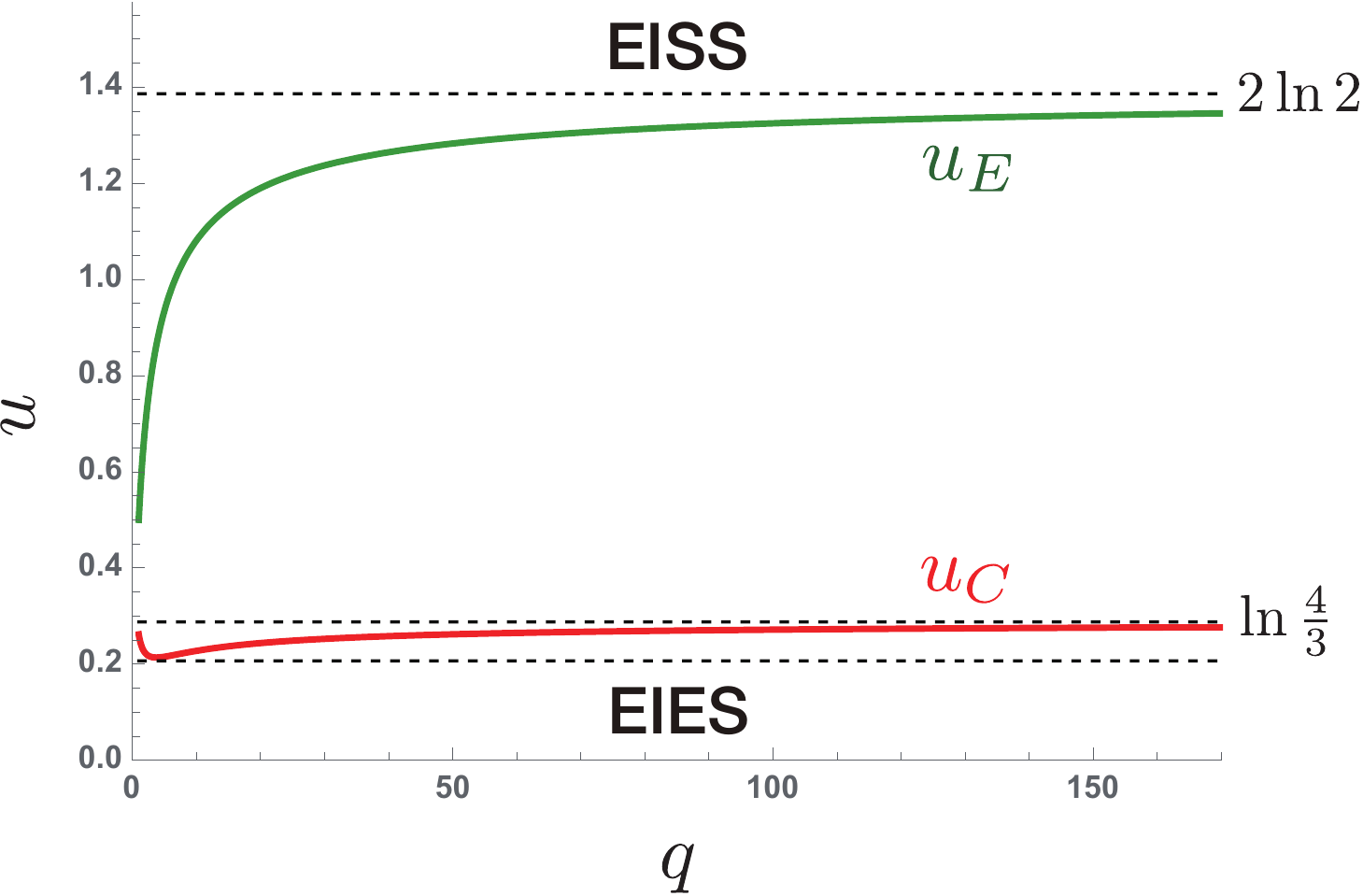}
\caption{
(Color online) Left. Phase diagram of the entanglement spectrum, ($q,u$)-plane. The entanglement $S_{q}= \ln N - u$ decreases as $u$ increases. 
The entangled phase (red) is for $0 < u < u_C$: the eigenvalue distribution is a (deformed) semicircle around $1/N$; as $u\downarrow0$ the semicircle degenerates into a Dirac delta and the state is maximally entangled.
The critical line $u=u_C(q)$ is the ``concentration" line (pushed-to-pulled transition) where the gap closes, the left endpoint of the entanglement spectrum touches the boundary $\lambda=0$ and the entanglement spectrum starts developing a sharp concentration of eigenvalues near zero; $u_C(q) \to  \ln (4/3)$ as $q\to\infty$.
The typical phase (blue) is for $u_C < u < u_E$: 
the critical line $u=u_E(q)$ is the ``evaporation" line, corresponding to typical states. The largest eigenvalue evaporates from the spectrum sea, characterized by a Mar\v{c}enko-Pastur distribution; $u_E(q) \to 2 \ln 2$ as $q\to\infty$.
The separable phase (green) is for $u_E< u< \ln N$: the line $u= \ln N$ (not shown) corresponds to separable states.
Right. Same phase diagram for a wider range of $q$. Observe that $u_C(q)$ has a minimum $u_C\simeq 0.214$ at $q\simeq 3.733$. States EIES (entropy-independent entangled states) below the lowest dashed line are entangled independently of the adopted entropy measure (namely, the value of $q$). States EISS (entropy-independent separable states) above the dashed asymptote have a significant ($O(1)$)  separable component independently of the adopted entropy measure (namely, the value of $q$).
}
\label{fig:phasediagram}
\end{figure}

As anticipated, as $q$ and $u$ are varied, one encounters two critical lines. See Fig.~\ref{fig:phasediagram}.
Starting from small values of $u$ (large entanglement $S_{q}= \ln N - u$ across the bipartition), one encounters a first phase transition at $u=u_C(q)$.
This first critical line separates an ``entangled" phase (red in the figure), present for $0< u< u_C$, from a ``typical" phase (blue in the figure).  

In the entangled phase the entanglement spectrum is a (deformed) semicircle around $1/N$. Notice that as $u\downarrow 0$ the semicircle degenerates into a Dirac delta, corresponding to maximally entangled states.

Across the critical line $u=u_C(q)$  the so-called pushed-to-pulled transition takes place. 
We call it the ``concentration" line, here the gap closes and the entanglement spectrum touches the boundary $\lambda=0$, so that the left endpoint $a=0$. Above this critical line a sharp concentration of eigenvalues near zero is formed with the development of a sharp (integrable) spike. Observe that $u_C(q) \to  \ln (4/3)$ as $q\to\infty$. Along the  concentration line $u=u_C(q)$ the phase transition is third order~\cite{DPFPPS}, \emph{except} at $q=1$, where it becomes fourth order~\cite{vnentr}.

As $u$ increases, for $u_C < u < u_E$, one finds the typical phase. Eventually, one reaches the second critical line $u=u_E(q)$, which separates the typical phase from a ``separable" phase (green in the figure), present for $u_E< u< \ln N$. 

The value $u=u_E(q)$ is characterized by the onset of evaporation of the largest eigenvalue. For $u=u_E(q)$ the distribution is Mar\v{c}enko-Pastur. Observe that $u_E(q) \to 2 \ln 2$ as $q\to\infty$, and $u= \ln N$ for genuinely separable states. Along the evaporation  line $u=u_E(q)$ the phase transition is first order~\cite{DPFPPS}, \emph{except} at $q=1$, where it becomes second order~\cite{vnentr}.
Interestingly, at $q=1$ the phase transitions are softer.

All phase transitions are detected by a change in the entanglement spectrum. This is reflected in a sharp variation of the relative volume of the manifolds with constant entanglement (isoentropic manifolds)~\cite{DPFPPS}.

We observe that states above the first asymptote, at $q=2 \ln 2$, always have a significant separable component $O(1)$, independently of the value of $q$, and therefore of the R\'enyi entropy used to measure entanglement. We call them ``entropy-independent separable states" (EISS). 
States below the parallel line tangent to the minimum of $u_C(q)$
are significantly entangled, independently of the value of $q$, and therefore of the particular R\'enyi entropy. We call them ``entropy-independent entangled states'' (EIES). 
Further investigation is needed to understand what EISS and EIES are and if they have some sort of characterization.

\section{Conclusions and perspectives}
\label{sec:concl}

We have determined the phase diagram of the entanglement spectrum for a bipartite quantum system. 
The analysis hinges upon saddle point equations and a Coulomb gas method. It is valid in the limit of large quantum systems. 

The present analysis basically completes the characterization of typical bipartite entanglement.
\emph{Multipartite} entanglement is much more difficult to study. One possible strategy consists in looking at the distribution of bipartite entanglement when the bipartition is varied~\cite{FFPP,FFMPP}. This problem is more difficult, and no complete characterization exists, although a number of interesting ideas have been proposed, in particular for small subsystems~\cite{GALRZ,GZ,GBZ,GRDMZ,Wallach,guhne}. One of the main roadblocks seems to be the presence of frustration~\cite{frustr}, which makes the analysis (and the numerics) more involved. 
A thorough understanding of multipartite entanglement is however crucial, also in view of possible 
quantum applications~\cite{adesso}.


\ack
We thank Beppe Marmo for interesting discussions.
PF and SP are partially supported by Istituto Nazionale di Fisica Nucleare (INFN) through the project ``QUANTUM". PF is partially supported by the Italian National Group of Mathematical Physics (GNFM-INdAM). KY is partially supported by the Top Global University Project from the Ministry of Education, Culture, Sports, Science and Technology (MEXT), Japan, by the Grants-in-Aid for Scientific Research (C) (No.~18K03470) and for Fostering Joint International Research (B) (No.~18KK0073) both from the Japan Society for the Promotion of Science (JSPS), and by the Waseda University Grant for Special Research Projects (No.~2018K262).


\end{document}